\documentclass[letterpaper,12pt]{article}   %% LaTeX 2e (preferred)

\usepackage{osajnl2} %% do not use with REVTeX4
\usepackage{graphics, epsfig}

\begin{document}

\title{Cancellation of simple optical anisotropies without  use of a Faraday mirror}

%% For REVTeX it is possible to automate superscript and e-mail callouts with the superscriptaddress option; see REVTeX4 documentation.

\author{\bf Rajendra~Bhandari}
\address{  Raman Research Institute, \\ Bangalore 560 080, India. \\ email: bhandari@rri.res.in}

\begin{abstract} We first derive the round-trip 
Jones matrix for double passage through a reciprocal optical medium by means of reflection off a 
plane mirror that could be optically anisotropic. We then show that if a medium with only linear 
birefringence and linear dichroism is placed between a pair of orthogonal quarterwave plates  
with principal axes  at $45^\circ$ to that of the medium and the sandwich is placed in front of an 
isotropic mirror it behaves, under double passage, like an isotropic medium. We describe a  simple 
liquid crystal device that behaves, in reflection, as an isotropic medium whose refractive index 
can be varied by application of an electric field, thus acting as a phase only modulator for light 
in any polarization state.\\

\end{abstract}


\maketitle %% null function with osajnl.sty

When a monochromatic plane polarized  light wave changes its direction of propagation 
in space either by means of  reflection or by means of refraction, 
the description of the polarization state in the new direction involves the 
choice of a new set of basis states along with a new set of phases for 
these basis states. Definition of a round-trip Jones matrix for double 
passage through a medium requires a clear statement of these choices.  
A  natural choice is  the following: Take a set of 
linearly polarized vibrations in and perpendicular to a chosen plane of incidence 
(p and s states), say along $\hat x$ and $\hat y$, vibrating in phase, as the basis states
for the original direction of propagation. Rotate the two linear vibrations 
about an axis perpendicular to the 
plane of incidence by an angle equal to that between the final and initial 
directions so that $\hat x$ and $\hat y$ go to $\hat x'$ and $\hat y'$ respectively. A set of 
linear vibrations along $\hat x'$ and $\hat y'$ in phase with each other then provides 
a natural choice for  the basis states 
as well as their phases, for the new direction of propagation. We shall call this 
the ``travelling frame convention". This is the convention used to write 
the Fresnel's relations for reflection and refraction at an interface
for arbitrary angles of incidence 
\cite{bornwolf} and to define the reflection Jones matrix in ellipsometry. 
Considerations in \cite{decomposition} also support the above choice.
The reflection Jones matrix $Z$ acting on the initial Jones vector $\mid i>$ gives a 
new  vector $\mid f>$ whose components represent the amplitudes with respect 
to the new base states. 

Therefore let the  polarization basis states for the initial direction of propagation be chosen to be
(i) $~E_x=E{\rm exp}(i\omega t),~ E_y=0 ~$ and (ii)$~ E_x=0, ~E_y=E{\rm exp}(i\omega t)$,
where $E_x$ and $E_y$ are the $x$ and $y$ components of the electric field in the wave.
Let the initial Jones vector be $\mid i>$=col.$[c_1,c_2]$. The fields in the initial wave are 
then $E_x=c_1E{\rm exp}(i\omega t),~~ E_y=c_2E{\rm exp}(i\omega t)$. In the above stated convention 
the  basis states for the final direction of propagation are
(i)$~ E_{x'}=E{\rm exp}(i\omega t),~ E_{y'}=0 ~$ and (ii)$~ E_{x'}=0,~ E_{y'}=E{\rm exp}(i\omega t)$.
If $\mid f>=Z\mid i>$=col.$[c'_1,c'_2]$ is the 
final Jones vector, where $Z$ is a Jones matrix, the fields in the final wave  are given by 
$E_{x'}=c'_1E{\rm exp}(i\omega t),~~ E_{y'}=c'_2E{\rm exp}(i\omega t)$.

We next use the method described in \cite{decomposition,physicab} to  derive a round trip Jones matrix 
for double passage through a reciprocal optical medium $M$ by means of reflection off  
an optically anisotropic mirror (Fig.1(a)). In this Letter we shall ignore 
all constant isotropic phase factors equal to $i$ and $-1$. Let $\hat z$ and $-\hat z$ be  
the two directions of propagation  
and let the reflection plane be 
chosen to be the $(\hat x,\hat z)$ plane \cite{footnote}. Following \cite{decomposition,physicab}, 
the actual optical circuit shown in Fig.1(a) can be  replaced by an equivalent circuit
shown in Fig.1(b) where the mirror has been replaced by the reflection matrix $Z$ of
the  surface determined with the  above convention and the medium 
$M$ during reverse passage has been replaced by  $\bar M$, 
obtained from $M$ by rotation about $\hat y$ through $\pi$. It was shown in \cite{transpsymm}
that 

\begin{eqnarray}
\bar M =
\left( 
  \begin{array}{lr}
m_{11} & -m_{21} \\
-m_{12} &m_{22}\\
\end{array}
\right)~{\rm for}~ M=\left( 
  \begin{array}{lr}
m_{11} & m_{12} \\
m_{21} &m_{22}\\
\end{array}
\right) .& & \label{eq:m1}
\end{eqnarray}

\noindent As shown in Fig.1(b), the equivalent evolution consists of two parts: (a) 
a pure polarization evolution given by a product matrix $M^{(\rm rt)}$ 
that we shall call the round trip Jones matrix and (b) a $\pi$ rotation 
of the beam about $\hat y$. The operation (b) does not affect the 
ellipticity of the polarization ellipse but does change its absolute orientation 
in space. The round trip Jones matrix $M^{(\rm rt)}$
is therefore given by,  

\begin{eqnarray}
 & & M^{(\rm rt)}=  \bar M Z M . \label{eq:m2}
\end{eqnarray}

\begin{figure*}
\centerline{
\epsfxsize=0.7\textwidth
\epsfbox{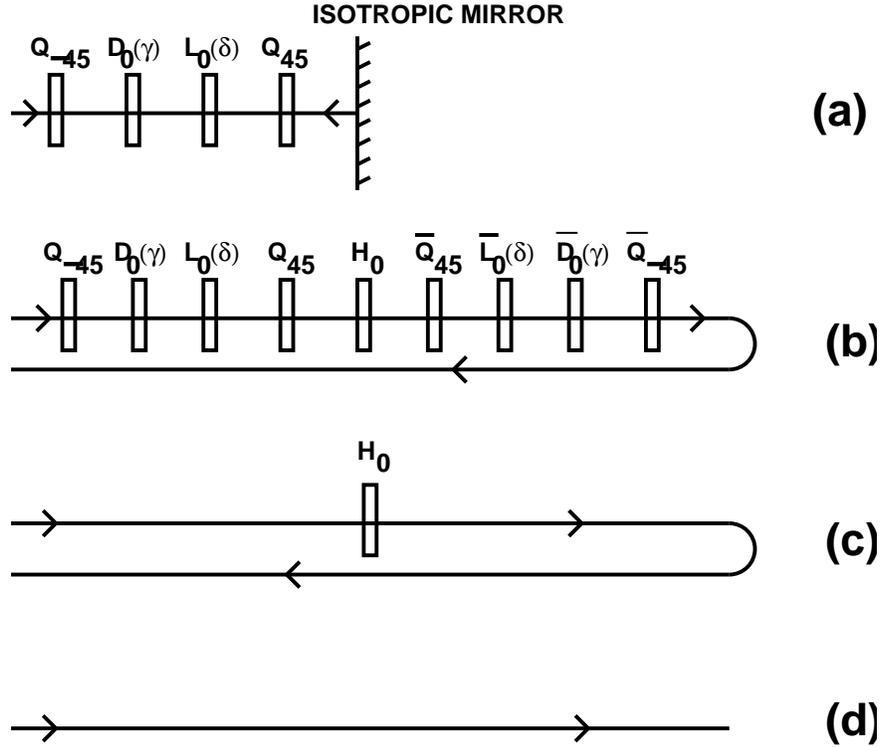}
}
\caption{A sandwich consisting of a reciprocal optical medium with  linear birefringence 
and linear dichroism along the same axis, placed between a pair of orthogonal 
quarterwave plates  at $+45^\circ$ or $-45^\circ$ behaves like an isotropic medium when placed in front of an 
isotropic mirror. 
(a) The actual optical configuration,  (b) the equivalent optical circuit,
(c) the reduced  circuit  and (d) an equivalent reduced  
circuit in the fixed frame. The isotropic factors  
$a$, $p$ and $r$  have been omitted for convenience.}
\label{fig.2}
\end{figure*}

\noindent If $\mid f>=M^{(\rm rt)}\mid i>={\rm col}.[c'_1,c'_2]$ is the final 
Jones vector, the fields resulting from operation (A) are  $E_{x}=c'_1E {\rm exp}(i\omega t),~ E_{y}=c'_2E{\rm exp}(i\omega t)$, 
while those after taking into account both operations (A) and (B) are 
$E_{x'}=c'_1E {\rm exp}(i\omega t),~ E_{y'}=c'_2E{\rm exp}(i\omega t)$, where $\hat x'=-\hat x ~{\rm and}~ \hat y'=\hat y$.

While this completes the description of our method, it is worthwhile 
to take note of another useful convention often used in the analysis of 
double passage problems. In this convention, which holds only for normal 
incidence, both the initial and 
the return Jones vectors are expressed in the same i.e. the initial set of basis 
states. In other words a given polarization ellipse in space is represented 
by the same Jones vector irrespective of whether the beam is travelling 
forward or backward \cite{jones1}. The Jones matrix describes the evolution of this 
ellipse.  We shall 
call this the ``fixed frame convention". 
Since the only change involved is the sign of the $x$-component of the electric field 
in the final wave, 
it is equivalent to  the action of 
a half-wave plate $H_0$ where $H_0$ is a half-wave plate with its fast axis 
along $\hat y$.
One can therefore describe the evolution of the polarization ellipse in 
the fixed frame as shown schematically in Fig.1(c) in terms of a matrix
$M'^{(\rm rt)}$ given by 

\begin{equation}
M'^{(\rm rt)}=H_0 \bar M Z M. \label{eq:m3} 
\end{equation}

When the mirror is optically isotropic, $Z=rH_0$
where $r$ is a constant isotropic attenuation factor representing the 
reflectivity of the surface \cite{footnote2}. 
For this case  Eqns. (\ref{eq:m2}) and (\ref{eq:m3}) are replaced by, 
$M^{(\rm rt)} =r\bar M H_0 M ~~{\rm and}~~ M'^{(\rm rt)}=rH_0 \bar M H_0 M$.

\begin{figure*}
\centerline{
\epsfxsize=0.7\textwidth
\epsfbox{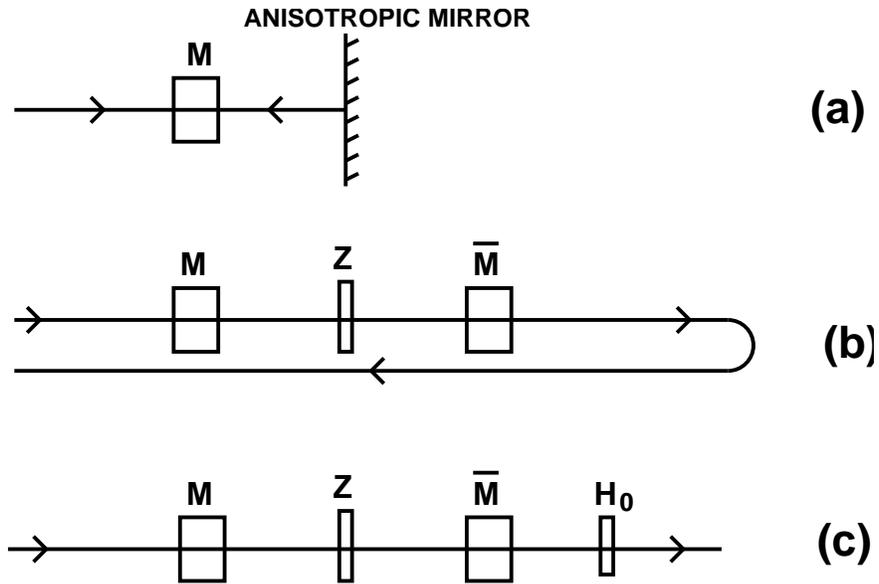}
}
\caption{(a) Double passage of a polarized light beam through a reciprocal medium by 
means of normal reflection off an optically anisotropic surface,
(b) the equivalent optical circuit consisting of a pure polarization 
evolution given by the Jones matrix $\bar M ZM$ followed by a rotation of the beam about an axis 
perpendicular to the plane of 
incidence and (c) an equivalent circuit in the fixed frame for the evolution shown in (b).}
\label{fig.1}
\end{figure*}

It is well known that double passage of a polarized light beam through a reciprocal medium 
by reflection off a Faraday mirror
can cancel an arbitrary optical anisotropy in the medium \cite{martinelli,vandeventer,transpsymm}. 
In the following we first show that  if a reciprocal  medium that has 
only linear birefringence or linear dichroism or both, with the same principal 
axes, is placed between a pair of orthogonal 
quarter-wave plates  with principal axes  at $+45^\circ$ or $-45^\circ$ to that of the medium 
and the sandwich is placed in front of an isotropic mirror as shown in Fig.2(a), it 
behaves, under double passage, like an isotropic medium placed in front of a mirror. 
The following matrix 
identities will be useful in proving the result.

\begin{eqnarray}
 & &  L_\phi(\delta) = R(\phi)L_0(\delta)R(-\phi),~D_\phi(\gamma) = R(\phi)D_0(\gamma)R(-\phi),\label{eq:i1}\\
& & {\bar L}_\phi(\delta)=L_{-\phi}(\delta),~{\bar D}_\phi(\gamma)=D_{-\phi}(\gamma),\label{eq:i2}\\
& & D_\phi(\gamma) D_{\phi+90^\circ}(\gamma)=L_\phi(\delta) L_{\phi+90^\circ}(\delta)={\bf 1},\label{eq:i3}\\
& & R(\phi_1)R(\phi_2)=R(\phi_1+\phi_2), R(\pm 180)=-{\bf 1}\label{eq:i4}\\
& & H_{\phi}=L_\phi(\pi),~Q_{\phi}=L_\phi(\pi/2), ~Q_\phi H_\phi Q_\phi = -{\bf 1}\label{eq:i5}\\
& & {\rm and}~~R(\phi)H_0R(\phi)=H_0, \label{eq:i6}
\end{eqnarray}

\noindent where $L_\phi(\delta)$ stands for a linear retarder with retardation $\delta$ and with
its fast axis making an angle $\phi$ with $\hat y$, $D_\phi(\gamma)$ is a 
linearly dichroic element with relative
attenuation coefficient $\gamma$ and its easy axis at an angle $\phi$ with $\hat y$ 
and $R(\phi)$ is an 
optical rotator with rotation $\phi$ in real space, all three being elements of the 
Special Linear group SL(2,C).
If the principal axes of the
medium are aligned to be along $\hat x$ and $\hat y$, the Jones matrix $M$ of the 
medium for single passage is given by,

\begin{eqnarray}
 & & M  = a~ p~  L_0(\delta)D_0(\gamma),   \label{eq:c1}
\end{eqnarray}

\noindent where $a$ and $p$ are the isotropic 
absorption and  phase factors for single passage. The equivalent optical circuit 
for double passage is shown in Fig.2(b). The round trip Jones matrix 
$M^{(\rm rt)}$ for the polarization evolution  is given by, 

\begin{eqnarray}
 & & M^{(\rm rt)}  = ra^2 p^2\bar Q_{-45}\bar D_0(\gamma)\bar L_0(\delta)\bar Q_{45}H_0  Q_{45} L_0(\delta) D_0(\gamma) Q_{-45}\nonumber\\
 & & = ra^2 p^2Q_{45}D_0(\gamma)L_0(\delta)Q_{-45}H_0  Q_{45} L_0(\delta)D_0(\gamma) Q_{-45} \nonumber\\
 & &  =ra^2 p^2 R(45)Q_0 R(-45)D_0(\gamma)L_0(\delta)R(-45)Q_0 R(45)H_0 \nonumber \\
 & & \times  R(45)Q_0 R(-45)L_0(\delta)D_0(\gamma)R(-45)Q_0 R(45) \nonumber\\
 & &  =-ra^2 p^2 R(45)Q_0 R(-45)D_0(\gamma)L_0(\delta)R(-90)L_0(\delta)D_0(\gamma)\nonumber\\
 & &  \times R(-45)Q_0 R(45) \nonumber\\
 & &  =-ra^2 p^2 R(45)Q_0 R(-45)R(-90)R(90)D_0(\gamma)L_0(\delta)R(-90) \nonumber\\
 & &  \times L_0(\delta)D_0(\gamma)R(-45)Q_0 R(45) \nonumber\\
 & &  =-ra^2 p^2 R(45)Q_0 R(-45)R(-90)D_{90}(\gamma)L_{90}(\delta)L_0(\delta)D_0(\gamma)\nonumber\\
 & &  \times R(-45)Q_0 R(45) \nonumber\\
 & &  =-ra^2 p^2 R(45)Q_0 R(-180)Q_0 R(45) \nonumber\\
 & & = ra^2 p^2 H_0. \label{eq:b10}
\end{eqnarray}

\noindent where  one or more of the  identities (\ref{eq:i1}) - (\ref{eq:i6}) have been used  
in each step of simplification. Fig.2(c) shows the final result.   $M'^{(\rm rt)}$  
is given by,

\begin{eqnarray}
 & & M'^{(\rm rt)}= H_0 M^{(\rm rt)}  = ra^2 p^2{\bf 1}. \label{eq:b11}
\end{eqnarray}
 
\noindent This is shown in Fig.2(d).  Equations.(\ref{eq:b10}) and (\ref{eq:b11}) together with Figs. 2(c) and 2(d) 
demonstrate the  equivalence of the device 
with an isotropic medium placed in front of an isotropic mirror (compare with Fig.1 with $M={\bf 1}, Z=H_0$). 
Since $M^{(\rm rt)}$ and $M'^{(\rm rt)}$ are independent of $\delta$ and $\gamma$, 
the anisotropies have effectively been cancelled. The isotropic factors 
$a$, $p$ and $r$ however survive. A similar result is obtained if $Q_{\pm45}$ is replaced by $Q_{\mp45}$.

Using a similar  analysis  it can be shown that if 
an optical element with circular dichroism and optical activity, which 
can be expressed as a product 
$Q_{\pm45}D_0(\gamma)Q_{\mp45}R(\phi)$, is placed in front of an isotropic mirror, 
the round trip matrix for double 
passage is the same as that of an isotropic medium placed in front of the mirror. 
Optical activity and circular dichroism can therefore be cancelled just by reflection off an isotropic 
mirror. We also point out that in the above two examples, since $M'^{(\rm rt)}={\bf 1}$, the polarization 
ellipse is unchanged in the fixed frame as in the case of an isotropic medium. In
the Faraday mirror cancellation, however, $M'^{(\rm rt)}=R(\pm90)$ and the polarization ellipse gets 
rotated through $90^\circ$.

We next present a practical application of  cancellation of linear birefringence.
A transparent linearly birefringent substance can be characterized by $\gamma=0$ (or $a=1$) and two refractive 
indices $n_e$ and $n_o$
representing the velocities of propagation for electric vectors along the two principal axes 
that can be aligned to be along  $\hat x$ and $\hat y$.  The 
isotropic phase factor $p$ is given by,
 
\begin{eqnarray}
 & & p={\rm exp}(i\phi_{iso}) ~{\rm with}~ \phi_{iso}=(2\pi d/\lambda)(n_e+n_o)/2, \label{eq:l1}
\end{eqnarray}

\noindent where $d$ is the thickness of the sample and $\lambda$ is the wavelength of 
light.  The retardation $\delta$ due to birefringence is given by
 
\begin{eqnarray}
 & & \delta=(2\pi d/\lambda)(n_e-n_o). \label{eq:l2}
 \end{eqnarray}

\noindent Any material in which $\phi_{iso}$ given by Eqn.(\ref{eq:l1}) can be 
varied using an electric field,  placed between two 
orthogonal quarterwave plates and the sandwich placed in front of a mirror as described above  acts as an isotropic medium 
with its refractive index adjustable by means of an electric field. An example of such a material is a 
suitably prepared nematic 
liquid crystal cell in which $n_e=<n_e(z)>$ varies with an applied electric field and $n_o$
remains constant \cite{chigrinov}. The two quarterwave plates could in fact form part of the liquid crystal cell itself. 
Such a device would act as a `phase only' modulator for light in any polarization state. 
A closely related device for modulating unpolarized light using  a nematic liquid crystal
cell and a $45^\circ$ quarterwave plate placed in front of a mirror has earlier been described  by 
Love \cite{love}. 

In this Letter we have ignored constant isotropic phase factors equal to $i$ and $-1$. 
This is justified in problems where one is interested only in phase changes. 
However
the applications described in  this Letter and in \cite{transpsymm} 
demonstrate that throwing away 
the  isotropic  factors altogether may at times be like  ``throwing away 
the baby with the bath". The device described in this Letter in fact works because
$\phi_{iso}$ can be  controlled using an electric field.
  We also note that the analysis  in this 
paper applies only in the regime of linear optics and 
not to double passage caused by optical phase conjugation.

Another possible 
application of the method described in this Letter is in the analysis of Michelson interferometers 
where the beam splitter and the retroreflecting mirrors  have optical 
anisotropy and polarization changing elements have been placed in one or 
both paths of the interferometer. In this application since one compares 
two beams travelling in the same direction, if the interferometer is planar, the 
second part of the evolution shown in Fig.1(b) can be ignored and the matrix 
$M^{(\rm rt)}$, computed separately for each of the two paths, provides a 
complete solution to the interference problem. The beam splitter reflection 
and the mirror reflection at normal incidence are treated similarly in 
this method, i.e. both are represented by their natural $Z$-matrix.

%% sample sizing command; other sizing commands (and graphics packages) may be used as well


\begin{thebibliography}{4pism}
\bibitem{footnote}
It can be shown that the final results of 
any calculation by this method is independent of the choice of the plane of incidence.
\bibitem{footnote2}
The factor $r$, though unimportant, should also have been present in the analysis 
described in \cite{transpsymm}.
\end{thebibliography}
\end{document}